# Visual Magnitude of the BlueWalker 3 Satellite

Anthony Mallama, Richard E. Cole, Scott Harrington and Paul D. Maley[1]

2022 November 28[2]

Contact: anthony.mallama@gmail.com

Abstract

Observations have been carried out in order to assess the optical characteristics of the BlueWalker 3 spacecraft which is the prototype for a new satellite constellation. The illumination phase function has been determined and evaluated. The average visual magnitude when seen overhead at the beginning or ending of astronomical twilight is found to be +1.4.

[2] This manuscript replaces the version dated 2022 November 14. A more robust illumination phase function has been determined from 146 magnitudes recorded through November 24.

1. Introduction

The prototype for a new constellation of artificial satellites, named BlueWalker 3 (BW3), was launched by the AST SpaceMobile company on September 11 of this year. This spacecraft was folded up on-board the launch vehicle and has now unfolded into a very large object. Magnitudes obtained after this deployment indicate that it has become extremely bright. Satellites like BW3 are beginning to interfere with observations being made by professional and amateur astronomers (Mroz et al. 2022, and Mallama and Young 2021).

This brief paper characterizes the visual brightness of BW3 based on observations recorded through November 24. Section 2 describes the apparent magnitudes that were measured by experienced visual observers. Section 3 addresses the illumination phase function which quantifies the brightness. Section 4 gives the characteristic magnitude and Section 5 presents the conclusions.

2. Observations

The magnitudes used in this study were obtained by the observers listed in Table 1. Most of the data were obtained with the unaided eye or through binoculars. Satellite magnitudes are measured by comparison with nearby reference stars of known brightness. This accounts for variations in sky transparency and brightness. The method is described in more detail by Mallama (2022). Some of the pre-deployment observations were derived from video recordings by Langbroek. He transformed the red-sensitive magnitudes to the V-band using an empirical formula based on the analysis of reference star measurements.

Table 1. Observer coordinates

| Observer | Latitude | Longitude | Ht(m) |
|---|---|---|---|
| R. Cole | 50.552 | -4.735 | 100 |
| K. Fetter | 44.606 | -75.691 | |
| S. Harrington | 36.062 | -91.688 | 185 |
| M. Langbroek | 52.154 | 4.491 | 0 |
| M. Langbroek | 52.139 | 4.499 | -2 |
| P. Maley | 33.811 | -111.952 | 654 |
| P. Maley | 32.857 | -113.220 | |
| P. Maley | 34.6 | 33.0 | 0 |
| A. Mallama | 38.982 | -76.763 | 43 |
| A. Mallama | 38.72 | -75.08 | 0 |
| R. McNaught | -32.27 | 149.16 | |
| J. Respler | 40.330 | -74.445 | |
| R. Swaney | 41.403 | -81.512 | |
| S. Tilley | 49.434 | -123.668 | 40 |
| S. Tilley | 49.418 | -123.642 | 1 |
| E. Visser | 53.109 | 6.108 | 46 |
| A. Worley | 41.474 | -81.519 | 351 |
| J. Worley | 41.474 | -81.519 | 351 |
| B. Young | 36.139 | -95.983 | 201 |
| B. Young | 35.831 | -96.141 | 330 |

Figure 1 illustrates the brightness increase of BW3 that was observed beginning on November 11 of this year. Before deployment the apparent magnitudes generally ranged from about 8 to 4, while afterward most span approximately 4 to 0. So, there is an increase of about 4 magnitudes or 40 times in brightness. These apparent magnitudes are affected by the distance from satellite to observer and phase angle. The next section describes how these selection effects are removed.

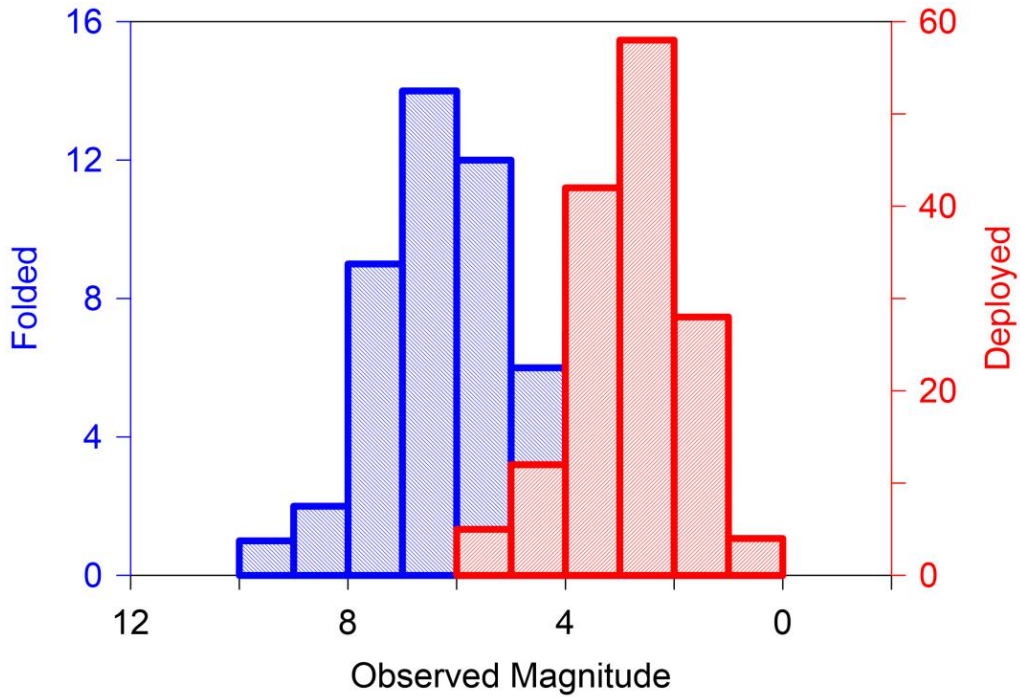

*Figure 1. Comparison of apparent magnitudes recorded before deployment with those taken after deployment and through November 24.*

3. Phase function

Satellite brightness may be characterized by determining the *phase function*. The parameter of characterization, called the phase angle, is that arc measured at the satellite between the directions to the Sun and to the observer. The phase angle varies during a pass as BW3 orbits the Earth.

Besides computing the phase angle, the apparent magnitudes are adjusted from the actual range to a standard distance of 1,000 km using the inverse square law of light. These distance-adjusted magnitudes are then fitted by least squares to the phase angles. The phase function resulting from magnitudes of BW3 obtained after deployment is illustrated in Figure 2. The coefficients of the quadratic polynomial fit are listed in Table 2.

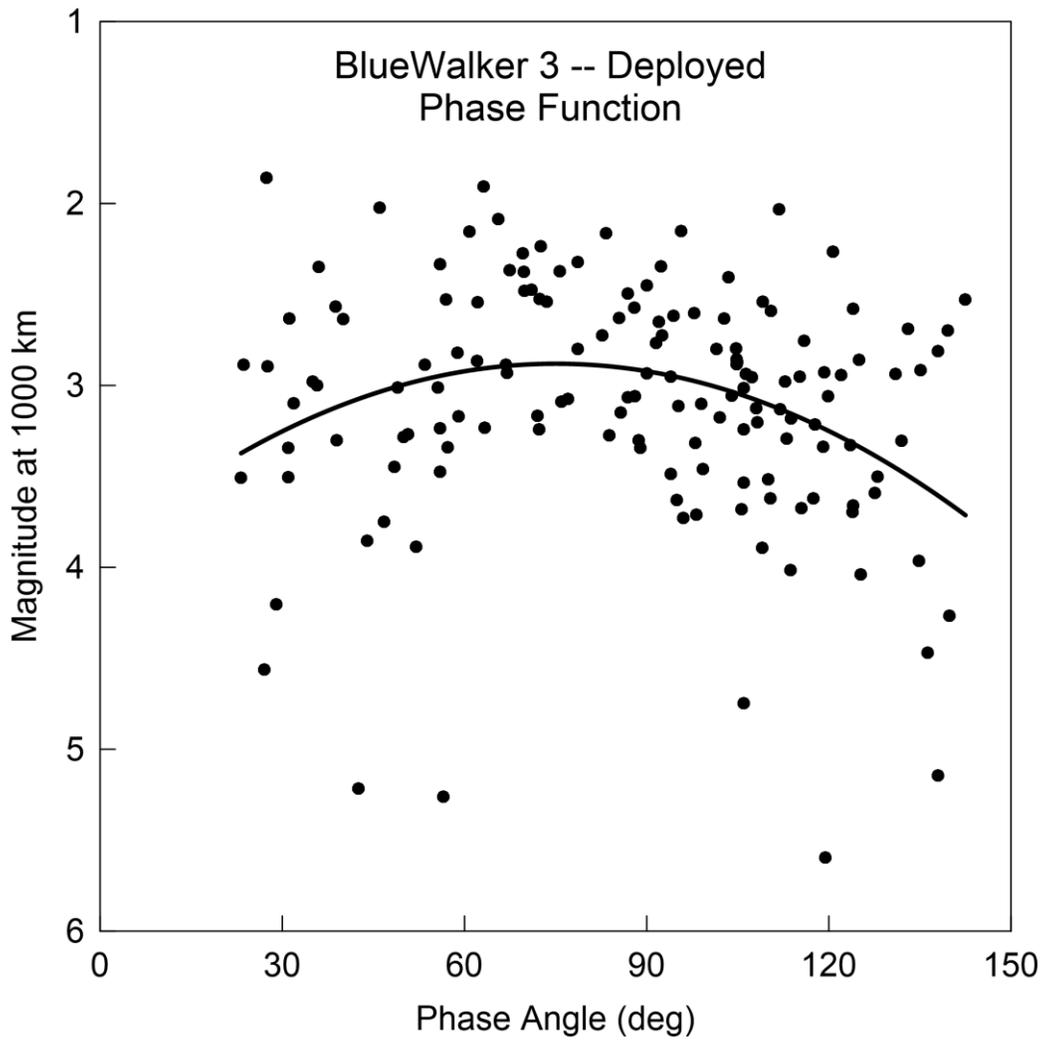

*Figure 2. The phase function determined from 146 magnitudes recorded after deployment and through November 24 of this year.*

Table 2. Phase function coefficients

```
Order    Coefficient
  0        3.912
  1       -0.02748
  2        0.0001831
```

4. Characteristic magnitude

A useful metric for observers, called the *characteristic magnitude,* is the average apparent brightness of a spacecraft when it is viewed overhead at the beginning or ending of astronomical twilight. Satellites interfere strongly with astronomical observations at this time because many of them are still illuminated by the Sun.

The phase angle corresponding to this geometry is 72$^o$, as illustrated in Figure 3, and the distance is adjusted to 510 km in the case of BW3 because that is its approximate altitude. When the phase function is evaluated for this angle and distance, the characteristic magnitude is +1.4.

It should be noted that satellites may be very bright when seen at other geometries, too. For example, specular polished surfaces can directly reflect sunlight to observers on the ground under a large variety of phase angles and slant ranges. Maley has seen the Soviet Molniya spacecraft at distances of 30,000 km with the unaided eye.

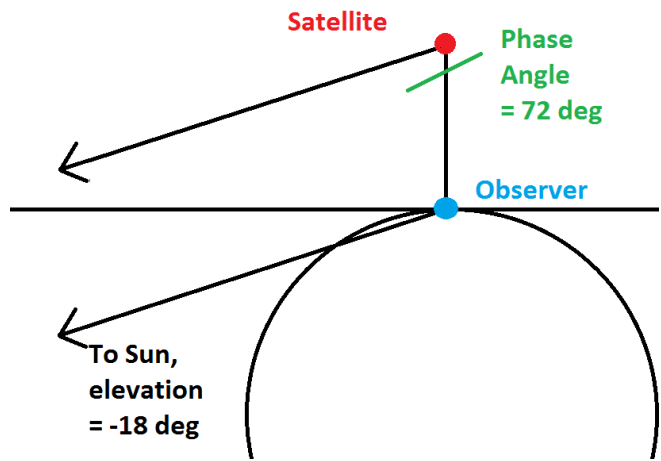

*Figure 3. The geometry of the characteristic magnitude is illustrated. Since the Sun is 18$^o$ below the horizon at the end of astronomical twilight, the phase angle for a satellite at zenith is 72$^o$.*

## 5. Conclusions

Visual magnitudes for the BW3 satellite obtained during the first 74 days after launch were analyzed. BW3 increased in brightness by 4 magnitudes or 40 times after the spacecraft deployed its large antenna. The characteristic magnitude of the unfolded spacecraft when seen overhead at the beginning or ending of astronomical twilight is +1.4.

## Acknowledgements

Scott Tilley provided information about the radio wavelength activity of BW3 which alerted us to its imminent deployment.

[1] Johnson Space Center Astronomical Society, Carefree, AZ, USA